\documentclass[aps,onecolumn,notitlepage,a4page,11pt]{revtex4-1} %preprintnumbers,floatfix,
\usepackage{amsmath,amssymb,bm}
\usepackage[]{graphicx}
\usepackage{subcaption}
\usepackage{soul}
\makeatletter
\usepackage[
       colorlinks=true,
      filecolor=black,
       anchorcolor=blue,
      linkcolor=blue,
      citecolor=cyan,%red,
      urlcolor=cyan,
       linktocpage=true,
        plainpages=false,
        breaklinks=true,
            pdfstartview=FitH
           ]{hyperref}

\def\obs{\text{obs}}
\def\th{\text{th}}

\def\H{\hat{H}}

\def\tta{ {t}}
\def\tc{\tau}

\def\tg{\tilde{g}}
\def\M{\mathcal{M}}

\def\({\left(}
\def\){\right)}
\def\[{\left[}
\def\]{\right]}

\begin{document}

%\preprint{}
%The View of 
\title{Observational Constraints on the Cosmology with Holographic Dark Fluid} 
% Dark Universe \Large Emergent Viewpoint in the Universe from %holographic the
%DGP type  Expanding 
% From Emergent Spacetime to  and Holographic Parameters
%Emergent universe on Holographic Cutoff and Parameter Constraints
\small
\author{Da Huang$^{1,2,3}$}
\email[Email: ]{dahuang@bao.ac.cn} 

\author{ Bum-Hoon Lee$^{4, 5}$}
\email[Email: ]{bhl@sogang.ac.kr} 

\author{Gansukh Tumurtushaa$^{5,6}$}
\email[Email: ]{gansuh@ntu.edu.tw} 
%%\email[Email: ]{gansuhmgl@ibs.re.kr}
%sichun@phys.ntu.edu.tw}

\author{Lu Yin$^{4,5}$}
\email[Email: ]{yinlu@sogang.ac.kr}
%\thanks{(corresponding author)}
%$^{\ddag}$
%\author{.}
%\email[Email: ]{cairg@itp.ac.cn} $^{1}$

%
\author{Yun-Long Zhang$^{1,2,3}$}  
\email[Email: ]{zhangyunlong@nao.cas.cn} 
%\thanks{(corresponding author)}
%zhangyunlong001@gmail.com}

%\author{...}
%\vspace{+25pt}
%\thanks{(corresponding author)}

%\affiliation{Center for Quantum Spacetime, Sogang University, Seoul 04107, Korea}
%\affiliation{Department of Physics, Sogang University, Seoul 04107, Korea}

\affiliation{$^1$ National Astronomy Observatories, Chinese Academy of Science, Beijing, 100101, China}

\affiliation{$^2$ School of Fundamental Physics and Mathematical Sciences, Hangzhou Institute for Advanced Study,
University of Chinese Academy of Sciences, Hangzhou 310024, China}

\affiliation{$^3$ International Center for Theoretical Physics Asia-Pacific, Beijing/Hangzhou, China}

%\affiliation{$^4$ Center for Theoretical Sciences, Sogang University} 
\affiliation{$^4$ Department of Physics, Sogang University, Seoul 04107, Korea}
\affiliation{$^5$ Center for Quantum Spacetime, Sogang University, Seoul 04107, Korea}

\affiliation{$^6$ Leung Center for Cosmology and Particle Astrophysics, National Taiwan University, Taipei 10617,
Taiwan} 

%%\affiliation{$^2$  APCTP }

%\affiliation{$^2$  IBS Center for Theoretical Physics of the Universe(CTPU),
%Institute for Basic Science(IBS), Daejeon 34051, Korea}

%%\affiliation{$^3$ Center for Gravitational Physics, Yukawa Institute for Theoretical Physics,
%%Kyoto University, Kyoto  606-8502,  Japan}
%Asia Pacific Center for Theoretical Physics, Pohang, Korea}
% 790-784,  and Center for Quantum Spacetime, Sogang University, Seoul 121-742, Korea}

\date{\today}

\begin{abstract}
\begin{center}  { \normalsize \bf Abstract} \end{center}
We consider the holographic Friedman-Robertson-Walker (hFRW) universe on the 4-dimensional membrane embedded in the 5-dimensional bulk spacetime, and fit the parameters with the observational data. In order to fully account for the phenomenology of this scenario, we consider the models with the brane cosmological constant and the negative bulk cosmological constant. The contribution from the bulk is represented as the holographic dark fluid on the membrane. We derive the universal modified Friedmann equation by including all of these effects in both braneworld and holographic cutoff approaches. For three specific models, namely, the pure hFRW model, the one with the brane cosmological constant, and the one with the negative bulk cosmological constant, we compare the model predictions with the observations. The parameters in the considered hFRW models are constrained with observational data. In particular, it is shown that the model with the brane cosmological constant can fit data as well as the standard $\Lambda$CDM universe. We also find that the $\sigma_8$ tension observed in different large-structure experiments can be effectively relaxed in this holographic scenario.

\vspace{20pt}
%\begin{center} as a result 
%{\it %\footnotesize Interestingly
%Essay written for the Gravity Research Foundation 2019 Awards for Essays on Gravitation}\\
%~~\\
%Submitted on April 1, 2019
%\end{center}
\end{abstract}
\date{\today} %April 1, 2019}
%\date{April 1, 2018}%December 27, 2017}  \today
%[submission Date: ]
%\pacs{}

\maketitle

\newpage
\tableofcontents

\allowdisplaybreaks
%\newpage
%$\sigma_8$, $H_0$ and $\Omega_m$ Similar discordance is found in the value of . 
%one of the most 

\allowdisplaybreaks

\section{Introduction}

The origin and properties of dark energy and dark matter are mysterious in modern cosmology. By treating dark energy as the cosmological constant and dark matter as the collision-less particles, the standard Lambda Cold Dark Matter ($\Lambda$CDM) model can explain the cosmic evolution and the observed large-scale structure very well. However, more and more severe tensions between local measurements and the Planck's observations~\cite{Abbott:2017wau} of the Hubble constant might indicate the new physics beyond the $\Lambda$CDM model. Thus, alternative well-motivated models of the dark sector should be carefully examined. One theoretically well-motivated scenario is the Dvali-Gabadadze-Porati (DGP)-type braneworld model, where the 4-dimensional Friedmann-Lemaitre-Robertson-Walker (FLRW) universe is consistently embedded into a 5-dimensional {\it flat} bulk spacetime~\cite{Dvali:2000hr,Deffayet:2001pu, Lue:2005ya}. In the spirit of the membrane paradigm \cite{Price:1986yy, Parikh:1997ma},  the utility of the Brown-York stress-energy tensor is the usual choice \cite{Brown:1992br}.
It has been shown that the extra-dimensional effects can be effectively encoded by the holographic Brown-York stress-energy tensor on the brane \cite{Cai:2017asf,Cai:2018ebs}. Note that the application of the Brown-York stress-energy tensor has been inspired by the Wilsonian renormalization group (RG) flow approaches of fluid/gravity duality~\cite{Bredberg:2010ky, Bredberg:2011jq, Cai:2011xv,Brattan:2011my,Cai:2012vr,Bai:2012ci,Cai:2012mg}, where the holographic stress-energy tensor on the holographic cutoff surface could be identified with one of the dual fluid directly. The concept of the cutoff fluid was then generalized into a flat bulk with the cutoff surface of the induced de Sitter and FLRW universe \cite{Cai:2017asf}.  Therefore, an effective description of the accelerated universe driven by the dark fluid could be investigated.

One may ask, is there a bulk action in the (4+1)-dimension from which the modified Friedman equation can be consistently derived? The answer is yes, and the action is similar to that in the DGP braneworld model~\cite{Dvali:2000hr}, which has been proposed as the holographic Friedman-Robertson-Walker (hFRW) model and been carefully studied in Ref.~\cite{Cai:2017asf,Cai:2018ebs}. However, compared with the old braneworld scenario, the motivation there was a little different in that it was inspired by the aforementioned fluid/gravity duality on the finite cutoff surface~\cite{Bredberg:2010ky, Bredberg:2011jq} (see Refs.~\cite{Cai:2011xv,Brattan:2011my,Cai:2012vr,Bai:2012ci,Cai:2012mg} for more generalization and application of the cutoff approach to the fluid/gravity duality). More interestingly, Ref.~\cite{Cai:2017asf} considered the possibility that the whole dark sector, including dark matter and dark energy, could have their origin from the holographic dark fluid in the extra dimension. Such a scenario was further supported by the cosmological observational data by the Markov-Chain Monte Carlo (MCMC) sampling analysis in which the dataset of the Type-Ia supernovae (SNIa)~\cite{Cai:2018ebs} and the direct measurement of the Hubble constant $H_0$ were employed. As a result, this model proposed a possible origin of the dark sector, and shed new light on the underlying structure of our universe. 
More importantly, an intriguing duality emerges between the two viewpoints on the dark matter: on the one hand, there is one extra dimension from the 5d bulk perspective; one the other hand, the emergent dark matter on the 4d brane can be described by the holographic stress-energy tensor, which could only talk with the Standard Model matter via the gravitational interaction.

In the present paper, we shall revisit the hFRW scenario in light of the recently discovered $H_0$ and $\sigma_8$ tensions. Moreover, we shall study some extended versions of the hFRW model by including the cosmological constant on the 4d brane or that in the 5d bulk so as to fully access the viability of this hFRW scenario. In our analysis, we first derive the universal modified Friedman equation that can be applied in all of these cases. Note that, in our derivation, we use two different methods to obtain this universal Friedman equation to have a better understanding of its physical meaning. A parameter of integration in one formulation is especially shown to be related to the bulk black brane mass in the other derivation. Then, by using the widely-accepted {\tt CosmoMC} \cite{Lewis:2002ah} package, we perform the MCMC sampling analysis to fit the models and obtain the best-fit values of model parameters. Here we shall focus on three concrete models: the pure hFRW model already studied in Ref.~\cite{Cai:2018ebs}, the hFRW model with a 4d positive cosmological constant on the brane, and the one with the 5d negative cosmological constant in bulk, respectively. In our work, besides the data from Type-Ia supernovae~\cite{Riess:2016jrr} and the direct measurement of $H_0$, we also include the latest CMB temperature fluctuation and polarization data from the Planck Collaboration and the Baryon Acoustic Oscillation (BAO) data from 6dF Galaxy Survey~\cite{Beutler:2011hx} and the Sloan Digital Sky Survey~\cite{Ross:2014qpa,Alam:2016hwk} in our dataset due to their extreme precision and constraining power on model parameters. Based on our fitting results, we also hope to shed some light on the long-standing problems of cosmology in the hFRW framework, such as the $H_0$ tension and the $\sigma_8$ tension. 

The paper is organized as follows. In Sec.~\ref{Sec_Bulk}, we derive the universal Friedman equation in the hFRW scenario, which can be applied to the aforementioned three concrete models. In Sec.~\ref{result}, we apply the {\tt CosmoMC}\cite{Lewis:2002ah} code to fit these three models. The fitting results, together with the best-fit parameters, are also presented in the same section. Finally, we conclude and make discussions in Sec.~\ref{Sec_Conclusion}. Some details for the derivation of the universal Friedman equation are delegated to the two Appendices, corresponding to the two different formulations of the hFRW scenario.

%%%%%%%%%%%%%%%%%%%%%%%%%%%%%%%%%%%%%%%%%%%%%%%%%%%%%%%%%%%%%%%%%%%%%%%%%%%%%%%%%%%%%%%%%%%%%%%%%%%%%%%%%%%%%%%%%%%%%%%%%%%%%%%%%%%%%%%%%%%%%%%%%%%%%%%%%%%%%%%%%%%%%%%%%%
\section{Holographic Dark Fluid on the Membrane}%{Three different approaches}
\label{Sec_Bulk}

%\section{AdS/DGP Type Universe}

%\begin{center}
%{\bf II. {FRW} Hypersurface in a Flat Bulk}
%\end{center}
We consider a  $(3+1)$-dimensional time-like hypersurface $\Sigma$ with metric $g_{\mu\nu}$, which is embedded in the $(4+1)$-dimensional bulk $\M$ with metric $\tg_{AB}$. %action ${\mS}_{4}$ and 
The total action is given by %${\mS}_{tot}={\mS}_{5}+{\mS}_{4}$, where % $$ action ${\mS}_{5}$ and
\begin{eqnarray}
{\cal S}_{\rm tot} &=& \int_{\cal M} d^5 x \sqrt{-\tilde{g}}\left[\frac{1}{2\kappa_5}(R_5-2\Lambda_5)\right] + \frac{1}{\kappa_5}\int_\Sigma d^4 x \sqrt{-g} {\cal K} + {\cal S}_4\,,\label{LagBulk}\\
{\cal S}_4 &\equiv& \int_\Sigma d^4 x \sqrt{-g} \left[\frac{1}{2\kappa_4}  \left(R_4-2\Lambda_4\right) +{\cal L}_M\right] \,,\label{LagBrane}
\end{eqnarray}
where ${\cal L}_M$ denotes the Lagrangian density for matter on the brane, including the radiation, baryon matter and dark matter. %, as well as possibly the dynamical dark energy. 
Since we are considering the evolution of our universe, we assume the 4d geometry on the brane is homogeneous and isotropic, which is usually described by the spatially flat FRW metric as follows:
\begin{eqnarray}
ds_4^2 = g_{\mu\nu} dx^\mu dx^\nu = -dt^2 + a(t)^2 dx_i^2\,,
\end{eqnarray}  
with $i=1,2,3$. By taking the functional derivative with respect to the bulk metric $\tilde{g}_{AB}$, we can obtain the following $(4+1)$-dimensional bulk Einstein equations,
\begin{equation}\label{EinsteinBulk}
 G_{MN} + \Lambda_5 \tilde{g}_{MN}=0.
\end{equation}
In the following, we shall use two different methods to derive the modified Friedman equation in the hFRW scenario, which can provide us two complementary viewpoints towards the origin of the dark sector in this geometric setup. 
 
% with a Cosmological Constant
\subsection{DGP-type universe embedded in the AdS bulk}\label{SecBrane}
% on 4d Freedmann Equation on the Brane
Based on the symmetry requirement, we take the bulk metric $\tilde{g}_{AB}$ to be in the Gaussian normal coordinate
\begin{eqnarray}\label{Metric5D}
d\tilde{s}^2_5 = -N(w,t)^2 dt^2 + A(w,t)^2 a(t)^2 dx_i^2 +dw^2\,,
\end{eqnarray}
and view our universe as a hypersurface $\Sigma$ located at a fixed position, say $w=0$, in this bulk spacetime, where $N(w,t)$ and $A(w,t)$ are two functions of the time variable $t$ and the extra dimension coordinate $w$. The boundary condition can be taken as follows: 
%with the boundary condition as follows
\begin{eqnarray}\label{bc0}
N(w,t)^2 = 1\,,\quad A(w,t)^2 = 1\,,\quad \mbox{when } w\to 0\,.
\end{eqnarray}
An important consequence of these boundary conditions is 
\begin{eqnarray}
\dot{A}(w,t) = 0\,, \quad \mbox{at } w=0\,, 
\end{eqnarray}
where and in the following the prime and dot denote the derivatives with respect to $w$ and $t$, respectively. We can obtain these two functions by solving the bulk Einstein equations in Eq.~(\ref{EinsteinBulk}) with the metric in Eq.~(\ref{Metric5D}), while taking into account the boundary conditions in Eq.~(\ref{bc0}). Consequently, the most general solution is given by
\begin{eqnarray}\label{Sol5D}
N &=& \frac{\partial_t [aA]}{\dot{a}} = A + \dot{A}/H\,, \nonumber\\
A^2 &=& \cosh(\lambda_5 w) - \frac{2H^2}{\lambda_5^2} \left[1-\cosh(\lambda_5 w)\right] + \sigma^\prime \frac{2\sinh(\lambda_5 w)}{\lambda_5} \sqrt{\frac{\lambda_5^2}{4} + H^2 + \frac{I}{a^4}}\,,
\end{eqnarray}
where $H \equiv \dot{a}/a$ is the Hubble parameter and $I$ is a constant of integration with $\lambda_5 \equiv -2\Lambda_5/3$ and $\sigma^\prime =\pm 1$. For convenience, we present the final solution here and delegate the details of our calculation in Appendix~\ref{App1}, which matches with the derivation in Refs. \cite{Binetruy:1999ut,Shiromizu:1999wj,Mukohyama:1999qx,Dick:2001sc}.

After specifying the bulk geometry, we can now determine the cosmology on the brane at $w=0$. With the Lagrangian in Eqs.~(\ref{LagBrane}), the Einstein equations on the 4d brane can be written by
\begin{eqnarray}
G_{\mu\nu} + \Lambda_4 g_{\mu\nu} = \kappa_4 \left( T^M_{\mu\nu} + T^D_{\mu\nu} \right)\,,
\end{eqnarray}
where 
\begin{eqnarray}\label{MatterEM}
T^M_{\mu\nu} &\equiv & -\frac{2}{\sqrt{-g}} \frac{\delta \left(\int_\Sigma d^4 x \sqrt{-g} {\cal L}_M  \right)}{\delta g^{\mu\nu}}\,,
\end{eqnarray}
denotes the stress-energy tensor of matter fields on the brane, while the bulk geometry induces the following geometric energy-momentum tensor:
\begin{eqnarray}\label{GeoEM}
T^D_{\mu\nu} &=& {\cal T}^{+}_{\mu\nu} - {\cal T}^-_{\mu\nu}\,,
\end{eqnarray} 
in which
\begin{eqnarray}
{\cal T}^\pm_{\mu\nu} \equiv \frac{1}{\kappa_5}  \left(g_{\mu\nu}{\cal K}^\pm -{\cal K}^\pm_{\mu\nu}  \right)\,,
\end{eqnarray}
where ${\cal K}^\pm_{\mu\nu} \equiv \tilde{g}^{M}_\mu \tilde{g}^N_\nu \nabla_{(M} n^{\pm}_{N)}$ stands for the extrinsic curvature of the brane $\Sigma$ associated with its embeddings in the bulk spacetime. Note that $n^\pm$ are both unit normal vectors pointing to the positive $w$, but approach the brane from $w=0^+$ and $w=0^-$. As usual, we impose a $Z_2$ symmetry under which the transformation of $w\to-w$ gives the same geometry. Concretely, we have two choices for the sign of $\sigma^\prime$ in Eq.~(\ref{Sol5D}):
\begin{eqnarray}\label{BulkSign}
\sigma^\prime =  \left\{ \begin{array}{cc}
1\,, & w>0 \\
-1\,, & w<0 \\
\end{array} \right. \quad  \mbox{or} \quad
\sigma^\prime =  \left\{ \begin{array}{cc}
-1\,, & w>0 \\
1\,, & w<0\,. \\
\end{array}
\right.
\end{eqnarray}
 Therefore, we can expect ${\cal K}^-_{\mu\nu} = - {\cal K}^+_{\mu\nu}$, ${\cal T}^{-}_{\mu\nu} = - {\cal T}^{+}_{\mu\nu}$ and, thus, $T^D_{\mu\nu} = 2 {\cal T}^+_{\mu \nu}$ for both cases. 
%with ${\cal K}^\pm_{\mu\nu} \equiv \tilde{g}^{M}_\mu \tilde{g}^N_\nu \nabla_{(M} n^{\pm}_{N)}$ the extrinsic curvature for embedding the constant $w$ surface into the bulk with $n^{\pm\, M}$ as the unit normal vector along $\pm w$ direction. 
As a result, we just need to compute the extrinsic curvature of brane on the positive $w$, which is given by  
\begin{eqnarray}
{\cal K}^+_{tt} &=& -N N^\prime|_\Sigma = -\frac{1}{2}\lambda_5 \left(C_2 + \frac{\dot{C}_2}{H}\right)\,, \nonumber\\
{\cal K}^+_{ii} &=&  a^2 A A^\prime|_\Sigma = \frac{1}{2} \lambda_5 a^2 C_2\,, \nonumber\\
{\cal K}^+ &=& g^{tt}{\cal K}^+_{tt} + g^{ij} {\cal K}_{ij}^+ = \frac{\lambda_5}{2}\left(4C_2 + \frac{\dot{C}_2}{H}\right) \,.
\end{eqnarray}
Thus,
\begin{eqnarray}
{\cal T}^{+\,t}_{\quad\,\, t} &=& \frac{3\lambda_5}{2\kappa_5} C_2\,,\nonumber\\
{\cal T}^{+\,i}_{\quad\,\, j} &=& \frac{\lambda_5 }{2\kappa_5} \left(3C_2+\frac{\dot{C}_2}{H}\right)\delta^i_j\,.
\end{eqnarray}
On the other hand, in order to mimic the real matter, we can represent the geometric stress-energy tensor as a perfect fluid with $T^{D\,\mu}_{\quad\,\,\nu} \equiv {\rm diag}(-\rho_D, p_D, p_D, p_D)$. Thus, we have
\begin{eqnarray}
\rho_D &=& -\frac{3\lambda_5}{\kappa_5} C_2 = \sigma \frac{6}{\kappa_5}\sqrt{\frac{\lambda_5^2}{4} + H^2 + \frac{I}{a^4}} = \sigma \rho_c\sqrt{\Omega_\ell} \sqrt{\Omega_5+ {\H}^2 + \Omega_I(1+z)^4} \,,\nonumber \\
p_D &=& \frac{\lambda_5}{\kappa_5} \left(3C_2 + \frac{\dot{C}_2}{H}\right) = -\rho_D - \frac{\dot{\rho}_D}{3H} = -\rho_D + \frac{1+z}{3} \frac{\rho_D}{dz}\,,
\end{eqnarray}
where the sign $\sigma$ takes the value of $-1(+1)$ for the first (second) solution in Eq.~(\ref{BulkSign}), respectively. The $z$ is defined as the cosmological redshift with $a(t) = 1/(1+z)$. Here we have defined $\Omega_I\equiv I/H_0^2$, $\Omega_5 \equiv \lambda_5^2/(4H^2_0)$, and $\sqrt{\Omega_\ell} \equiv 2\kappa_4/(\kappa_5 H_0)$, respectively. Therefore, the 4d Freedmann equation is simply given by
\begin{eqnarray}
H^2 =\frac{\Lambda_4}{3}+ \frac{\kappa_4}{3} (\rho_M+\rho_D)\,,
\end{eqnarray}
which can be further simplified to
\begin{eqnarray}\label{E2Eq}
{\H}^2(z) = \Omega_4+\Omega_M(z) +\sigma \sqrt{\Omega_\ell} \sqrt{\Omega_5+{\H}^2(z)+\Omega_I(1+z)^4}\,,
\end{eqnarray}
where ${\H}(z)\equiv {\H}(z)/H_0$ and $\Omega_4\equiv \frac{\Lambda_4}{3H_0^2}$.
Then ${\H}^2$ can be solved to
\begin{eqnarray}\label{E2Sol}
{\H}^2(z) =\Omega_4+ \Omega_M(z) + \frac{\Omega_\ell}{2} +\sigma \frac{\sqrt{\Omega_\ell}}{2} \sqrt{\Omega_\ell + 4 [\Omega_5+\Omega_4+ \Omega_M(z)+ \Omega_I (1+z)^4]}\,,
\end{eqnarray}
where the sign $\sigma$ dependence can be seen by comparing Eqs.~(\ref{E2Sol}) and (\ref{E2Eq}).
 As mentioned before, $\Omega_M$ contains all of the standard matter components in the cosmology. Therefore, $\Omega_M = \Omega_b + \Omega_\gamma + \Omega_\nu + \Omega_c  $ in which the terms from left to right correspond to the fractional densities of baryons, photons, neutrinos, cold dark matter, and the possible dark energy, respectively.   

\subsection{Moving brane on the holographic cutoff} 
\label{SecMovingBrane}

In this subsection, we shall provide the other perspective towards the modified Friedman equations given in Eq.~(\ref{E2Sol}). %give a further derivation of the above Freedmann equations on the brane. 
Interestingly, this time we can show the physical meaning of the integration constant $\Omega_I$ obtained by the previous methods. Our starting point is still the bulk and brane Lagrangians in Eqs.~(\ref{LagBulk}) and (\ref{LagBrane}). Note that the general solution to the bulk Einstein equation in Eq.~(\ref{EinsteinBulk}) can be shown to be given by
\begin{eqnarray}\label{MetricBulk}
ds_5^2 = -f(r) d {\tc}^2 + \frac{1}{f(r)} dr^2 + \frac{r^2}{L^2} dx_i^2\,,
\end{eqnarray}
where $i=1,2,3$, and
\begin{eqnarray}\label{DefF}
f(r) = \frac{r^2}{L^2} - \frac{m}{r^2}\,,
\end{eqnarray}
in which the length scale $L$ can be related to the 5-dimensional cosmological constant via $\Lambda_5 = -6/L^2$, and $m$ refers to the  mass of black brane. The details of the derivation of this general bulk solution are given in Appendix~\ref{App2}.
%Note that by 5d Einstein equation, I can only obtain the solution in Eq.~(\ref{DefF}), in which the constant term $k$ in Yun-Long's note is absent.  Bulk

Now we would like to embed our universe brane into this bulk solution. Since we should have the usual FRW metric on the brane, we let $r({\tta}) = L a({\tta})$, so that $dr = L \dot{a} d{\tta}$. Furthermore, by requiring 
\begin{eqnarray}
-d{\tta}^2 = -f[r({\tta})] d{\tc}^2 + \frac{dr^2}{f[r({\tta})]}\,,
\end{eqnarray}
we can derive the following relation
\begin{eqnarray}
\frac{d{\tc}}{d{\tta}} = \frac{1}{f(r)} \sqrt{f(r)+{L^2 \dot{a}^2}}\,.
\end{eqnarray}
Thus, the unit tangent vector can be defined as 
\begin{eqnarray}
u^N = \left(\frac{1}{f(r)}\sqrt{f(r)+L^2 \dot{a}^2}, L\dot{a}, 0, 0, 0\right) \,,
\end{eqnarray}
with $u^N u_N = -1$. From which by orthogonality we can determine the unit normal vector $n$ as follows
\begin{eqnarray}\label{Normal}
n_N = \sigma \left(L\dot{a}, -\frac{1}{f(r)}\sqrt{f(r)+ L^2 \dot{a}^2}, 0, 0, 0\right)^T\,,
\end{eqnarray}
where we have inserted $\sigma$ to denote the direction of the normal vector. As discussed before, the Freedmann equation on the brane should take the following form:
\begin{eqnarray}\label{Freedmann1}
3H^2 = \Lambda_4+\kappa_4 (\rho_M + \rho_D)\,, 
\end{eqnarray}
where the holographic energy density can be determined by the brane extrinsic curvature as 
\begin{eqnarray}
{\cal K}^+_{\mu\nu} = -\frac{\kappa_5}{2}\left[(p_D+\rho_D)u_\mu u_\nu + \frac{\rho_D}{3}g_{\mu\nu}\right].
%= -\frac{\kappa_5}{6}{\rm diag} \left(3p_D+2\rho_D,\, a^2 \rho_D,\, a^2 \rho_D,\, a^2\rho_D\right),
\end{eqnarray} 
Note that the factor of $1/2$ arises due to our assumption that the spacetime is $Z_2$ symmetric under the mirror of brane. 
With the bulk metric in Eq.~(\ref{MetricBulk}) and the brane embedding specified by the unit normal vector $n_M$ given in Eq.~(\ref{Normal}), we can determine the nonzero component of the extrinsic curvature tensor
\begin{eqnarray}\label{ExtK3}
{\cal K}^+_{{\tta}{\tta}} = \frac{\sigma (f^\prime + 2 L \ddot{a}) }{2\sqrt{f+ L^2 \dot{a}^2}}\,,\quad 
{\cal K}^+_{ii}  = -\frac{\sigma a}{L} \sqrt{f+ L^2 \dot{a}^2}\,,
\end{eqnarray}
where a prime stands for the derivative with respect to the coordinate $r$. For the derivation details of the extrinsic curvature expression, please see Appendix~\ref{App2}. %Ext}. 

By comparing the two expressions above for ${\cal K}^+_{\mu\nu}$, we can easily obtain the geometric energy density as follows
\begin{eqnarray}\label{density}
\rho_D = \frac{6\sigma}{\kappa_5} \sqrt{\frac{f(a)}{L^2 a^2} + H^2}\,.
\end{eqnarray}
By putting the obtained $\rho_D$ into the Freedmann equation in Eq.~(\ref{Freedmann1}), we have
\begin{equation}\label{Freedmann2}
H^2 =  \frac{\Lambda_4}{3}+\frac{\kappa_4 \rho_M}{3} + \frac{2\kappa_4 \sigma}{\kappa_5} \sqrt{\frac{f(a)}{L^2 a^2} + H^2}\,.
\end{equation}
We can further yield the expression for the Hubble parameter $H$ from this equation which is given by
\begin{eqnarray}
{\H}^2 %&=& \Omega_M + \frac{{\Omega_\ell}}{2} - \frac{\sigma\sqrt{{\Omega_\ell}}}{2} \sqrt{{\Omega_\ell} + 4\left[\Omega_M + \frac{f}{L^2 a^2 H_0^2}\right]} \nonumber\\
&=&\Omega_4+ \Omega_M + \frac{{\Omega_\ell}}{2} + \frac{\sigma\sqrt{{\Omega_\ell}}}{2} \sqrt{{\Omega_\ell} + 4\left[\Omega_4+\Omega_M +\Omega_5 + \frac{\Omega_I}{a^4} \right]}\,,
\end{eqnarray}
where the notations are defined as
\begin{eqnarray}\label{notations}
\sqrt{{\Omega_\ell}} \equiv \frac{2\kappa_4}{\kappa_5 H_0^2}\,,\quad 
\Omega_4\equiv \frac{\Lambda_4}{3H_0^2},\quad 
\Omega_5\equiv \frac{1}{L^2 H_0^2} = -\frac{\Lambda_5}{6H_0^2}
 \,, \quad \Omega_I \equiv -\frac{m}{L^4 H_0^2}\,.
\end{eqnarray}
Note that the sign of the solution $\sigma$ can be derived from Eq.~(\ref{Freedmann2}). As promised at the beginning of this subsection, the parameter $\Omega_I$ given as a constant of integration in the first derivation is now closely related to the black brane mass $M$ in Eqs.~(\ref{MetricBulk}) and (\ref{DefF}). 

For the general hFRW models, we have two branches of solutions corresponding to two different sign $\sigma$ in Eq.~(\ref{E2Sol}). Note that our hFRW model, when setting $\Omega_I = 0$ and $\Omega_4 = \Omega_5 = 0$, would be reduced to the widely-studied DGP model. Therefore, we shall follow the convention of the DGP model in the literature to name these two branches: the solution with $\sigma = -1$ corresponds to the normal branch, while the other one with $\sigma = +1$ to the self-accelerating branch. In the following numerical fittings, we will only concentrate on the self-accelerating branch of the solution, which has a well physical motivated positive energy density of the dark fluid in \eqref{density}. 
For the normal branch, we would like to discuss it elsewhere in the future.

\section{Fitting Result with the Observational Data}\label{result}
%Numerical Analysis
%After the presentation in last 
In this section, we present our global fitting results based on the MCMC sampling method. We modified the  ${\bf CAMB}~$\cite{Lewis:1999bs} and {\bf CosmoMC} program~\cite{Lewis:2002ah} for the holographic FRW models and $\Lambda$CDM with Planck + BAO + HST + JLA. 
Beginning with the general holographic dark fluid model with the modified Friedmann equation in ~(\ref{E2Eq}), we consider the following three cases: (i) the original hFRW model where $\Omega_4 = \Omega_5 = 0$, (ii) the hFRW+$\Omega_4$ model where $\Omega_5 = 0$ while $\Omega_4$ is a free parameter, and (iii) the hFRW+$\Omega_5$ model where $\Omega_4 = 0$ while $\Omega_5$ is a free parameter.

We employ the dataset that includes those of the CMB temperature fluctuation from \textit{Planck 2018} with \texttt{Planck\_highl\_TTTEEE}, \texttt{Planck\_lowl\_TT}, polarization~\cite{Aghanim:2019ame,Akrami:2019izv,Aghanim:2018eyx,Akrami:2018odb,Aghanim:2018oex}, the Baryon Acoustic Oscillation (BAO) data from 6dF Galaxy Survey~\cite{Beutler:2011hx} and the Sloan Digital Sky Survey (SDSS)~\cite{Ross:2014qpa,Alam:2016hwk}, the Hubble parameter data point $H_{0}=74.03_{-1.42}^{+1.42}$ from Hubble Space Telescope (HST) observations~\cite{Riess:2019cxk}. In addition, we also add 740 type-Ia supernovae data points from the ``Joint Light curves Analysis'' (JLA)~\cite{Betoule:2014frx}. 

We combine the data by adding the $\chi^2$ of each dataset,
\begin{eqnarray}
\label{eq:chi}
{\chi^2}={\chi^2_{BAO}}+{\chi^2_{CMB}}+{\chi^2_{SN}}+{\chi^2_{H_0}}.
\end{eqnarray}
For the observation  of BAO, we use the comoving sound horizon $r_{s}(z_d)$, where $z_{d}$ is the redshift at the drag epoch~\cite{Percival2010}. The distance ratio is considered as $d_{z}\equiv r_{s}(z_{d})/D_{V}(z)$, where the $D_{V}(z)$  is the volume-averaged distance that is defined by \cite{Eisenstein2005}
\begin{eqnarray}
D_{V}(z)\equiv\left[(1+z)^{2}D_{A}^{2}(z)\frac{z}{H(z)}\right]^{1/3}\,.
\end{eqnarray}
Here, $D_{A}(z)$ is the proper angular diameter distance and is written as
\begin{eqnarray}
D_{A}(z)=\frac{1}{1+z}\int_{0}^{z}\frac{dz'}{H(z')}\,.
\end{eqnarray}
The $r_{s}(z)$ can also be given by	
\begin{eqnarray}
r_{s}(z)=\frac{1}{\sqrt{3}}\int_{0}^{1/(1+z)}\frac{da}{a^{2}H({\scriptstyle z'=\frac{1}{a}-1})\sqrt{1+(3\Omega_{b}^{0}/4\Omega_{\gamma}^{0})a}},\end{eqnarray}
where $\Omega_{b}^{0}$ and $\Omega_{\gamma}^{0}$ are the values of baryon and photon density parameters in the present universe, respectively.
As a result, we can write the $\chi^2$ value of BAO dataset  as	
\begin{eqnarray}
\label{eq:chibao}
{\chi^2_{BAO}}= \sum_{i=1}^n \frac{\left[{D_V^{\th}/r_s}(z_i) - {D_V^{\obs}/r_s}(z_i)\right]^2}{\sigma^2_i},
\end{eqnarray}
where the subscripts of ``th'' and ``obs'' denote the theoretical and observational values of the volume-averaged distance, respectively. The $n$ corresponds to the number of BAO dataset, and $\sigma_i$ is the errors of the data points.%, given by Table~\ref{tab:1}.
% ${D_V/r_s}_{th}(z_i)$ and ${D_V/r_s}_{obs}(z_i)$ 

The redshift distance to the decoupling epoch $z_{*}$ can affect the CMB result and give strong constraints on the model in the high redshift region around $z\sim 1100$.
The $\chi^{2}$ result of the CMB observational dataset is given by
\begin{eqnarray}
\chi_{CMB}^{2}=(x_{i,CMB}^{\th}-x_{i,CMB}^{\obs})(C_{CMB}^{-1})_{ij}(x_{j,CMB}^{\th}-x_{j,CMB}^{\obs}),\end{eqnarray}
where $C_{CMB}^{-1}$ is the inverse covariance matrix. Here, $x_{i,CMB}\equiv\left\{l_{A}(z_{*}),R(z_{*}),z_{*}\right\}$ is given in terms of the acoustic scale $l_{A}$ and shift parameter $R$, which are defined by
\begin{eqnarray}
l_{A}(z_{*})&\equiv&(1+z_{*})\frac{\pi D_{A}(z_{*})}{r_{S}(z_{*})},\\
R(z_{*})&\equiv&\sqrt{\Omega_{m}^{0}}H_{0}(1+z_{*})D_{A}(z_{*}),
\end{eqnarray}
respectively.	

%SN AND h0

For the type-Ia  Supernovae  dataset,  the distance modulus is
\begin{eqnarray}
\mu_{\obs}\equiv m_B^*-(M_B-\alpha \times X_1+\beta\times C).
\end{eqnarray}
The $m_B^*$ is the observed peak magnitude, $C$ replaces the color of a supernova at maximum brightness, $X_1$ means the time stretching of the light curve, and $M_B$ is the absolute magnitude.
The luminosity distance $d_L (z)$ from the distance modulus of each supernova is
\begin{eqnarray}
\mu_{\th}\equiv m -M = 5 \log[d_L (z)/\text{Mpc}]+25,
\end{eqnarray}
where the $m$ and $M$ are the apparent and absolute magnitude of the Supernovae. The $r(z)$ is the comoving distance while $d_L (z)=(1+z)r(z)$ represents the luminosity distance.
For the $\chi^2$ of the supernovae dataset with errors can be given as
\begin{eqnarray}
\chi_{SN}^{2}=(\mu_{i,\obs}-\mu_{i,\th}) C_{SN}^{-1}(\mu_{j,\obs}-\mu_{j,\th}),
\end{eqnarray}
where %$\mu_{obs}$ is the rest-frame peak B-band magnitude of the SN, and $m_mod$ is the predicted magnitude of a SN. 
$C_{SN}$ is the $N \times N $ covariance matrix of the Supernovae.  
In addition, the $\chi_{H_0}^2$ function can be given as	
\begin{eqnarray}
\label{eq:chih0}
{\chi^2_{H_0}}= \sum_{i=1}^n \frac{\left({H_0^{\th}} - {H_0^{\obs}}\right)^2}{\sigma^2_{\obs, i}},
\end{eqnarray}
where the $H_0^{\obs}$ is the center value from HST observation and $H_0^{\th}$ is the theoretical value from our model. $\sigma^2_{\obs}$ is the errors from the HST result.

Based on these $\chi^2$ estimations, now we consider the global fitting of the $\Lambda$CDM and hFRW models to the data. The prior values of the cosmological parameters are listed in Table~\ref{tab:3}. To ensure both the negative and the positive contributions of the $\Omega_{I} = \alpha \times \Omega_{\ell}$, we choose the $\alpha$ parameter to have values within the $-1\leq \alpha\leq 1$.%interval as is seen Table~\ref{tab:3} 

\begin{table}[ht!]
	\begin{center}
		\caption{ The priors for cosmological parameters.  }
		\begin{tabular}{|c||c|} \hline
			Parameter & Prior
			\\ \hline
			{Model parameter $\alpha$} & $-1 \leq \alpha \leq  1$
			\\ \hline
			{Model parameter} $\Omega_4$& $0 \leq \Omega_4 \leq  1$
			\\ \hline
			{Model parameter} $\Omega_5$& $0 \leq \Omega_5 \leq  1$
			\\ \hline
			Baryon density parameter& $0.5 \leq 100\Omega_bh^2 \leq 10$
			\\ \hline
			CDM density parameter & $0.1 \leq 100\Omega_ch^2 \leq 99$
			\\ \hline
			Optical depth & $0.01 \leq {\tau} \leq 0.8$
			\\ \hline
			Neutrino mass sum& $0 \leq \Sigma m_{\nu} \leq 2$~eV
			\\ \hline
			$\frac{\mathrm{Sound \ horizon}}{\mathrm{Angular \ diameter \ distance}}$  & $0.5 \leq 100 \theta_{MC} \leq 10$
			\\ \hline
			Scalar power spectrum amplitude & $2 \leq \ln \left( 10^{10} A_s \right) \leq 4$
			\\ \hline
			Spectral index & $0.8 \leq n_s \leq 1.2$
			\\ \hline
		\end{tabular}
		%\vskip 0.2in
		\label{tab:3}
	\end{center}
\end{table}

\begin{figure}[h!]
	\centering
	\includegraphics[width=0.70 \linewidth]{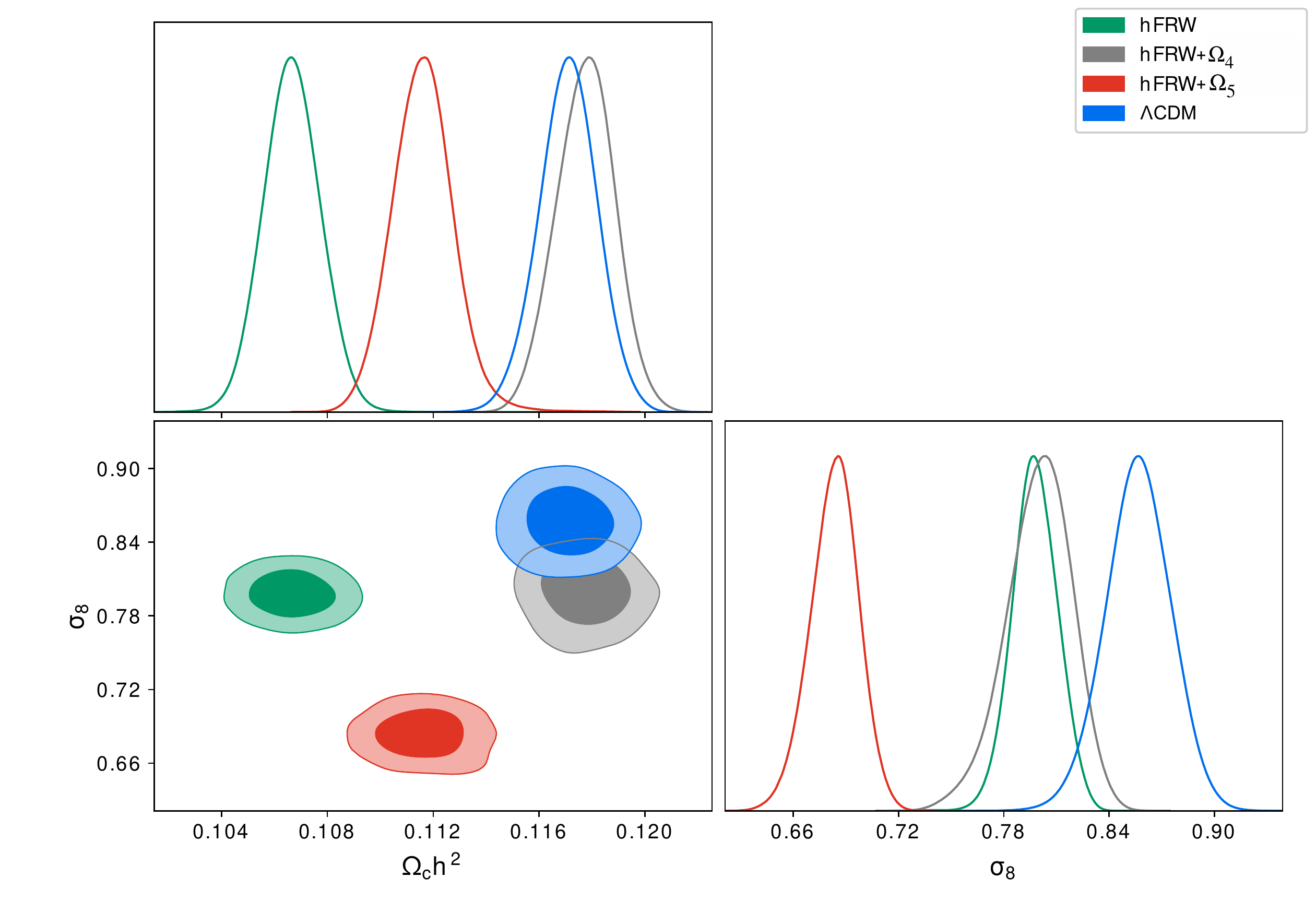}
\caption{\small
One and two-dimensional distributions of the $\Lambda$CDM, hFRW, hFRW$+\Omega_4$ and hFRW$+\Omega_5$ models using the dataset from CMB, BAO, Supernova and the $H_0$ value in HST. The contour lines represent 68$\%$(inner)~ and 95$\%$(outer)~ C.L., respectively. }% {\color{red} This fitting result under the dataset of CMB, BAO, Supernova and $H_0$ value in HST.}}
	\label{fg:6}
\end{figure}

The best-fit values of the cosmological parameters for each model are listed in Table~\ref{tab:2}. The table also summarizes the posterior distribution in Fig.~\ref{fg:6}. 
We found the $\chi^2$ minimum of $3102.49$ in the hFRW+$\Omega_4$ model, which is smaller than that of the  $\Lambda$CDM model, where the $\chi^2$ minimum value is 3019.66. The same best-fit values of hFRW and hFRW+$\Omega_5$ are obtained to be 3178.46 and 3251.54, respectively.

If the number of free parameters in different cosmological models is equal, one can employ the $\chi^2$ statistics to compare their statistical significance. In that case, a model with a smaller $\chi^2$ value is favored in describing data. However, when comparing the statistical significance of our models with the $\Lambda$CDM model, the $\chi^2$ value cannot make a fair comparison because, in general, models with more parameters have more tendency to have a lower value of $\chi^2$.  Although the $\chi^2$ value for hFRW+$\Omega_4$ is smaller than that of the $\Lambda$CDM model, such the result could be possible due to the additional $\{I, \Omega_4\}$ parameters of our model. Thus, to make a reasonable comparison, we apply the so-called Akaike Information Criterion (AIC)~\cite{Akaike1974}.  The AIC estimator is defined as $AIC \equiv - 2 \ln L_{max}+2k$, where $L_{max}$ and $k$ indicate the maximum likelihood and the number of free parameters, respectively. According to AIC and its interpretation, a smaller AIC value model is favored in describing data. Since we consider $\Lambda$CDM as a reference model, we use the pair difference value $\Delta AIC = AIC_{\Lambda\text{CDM}} - AIC_{\text{hFRW+}\Omega_4}$ between our model and $\Lambda$CDM model, which can also be rewritten as $\Delta AIC = \Delta\chi^2 - 2 \Delta k$. As a result, we obtain the relative difference of $\Delta AIC=3.17$, and the result suggests positive evidence against the $\Lambda$CDM model. If $\Delta AIC > 10$, which is the case for our remaining two models (hFRW and hFRW+$\Omega_5$), there is essentially no support concerning the reference model~\cite{Akaike1974}.
\begin{table}[h!]
 	\begin{center}
 		\caption{{\color{black}Fitting results at  68$\%$ C.L. for $\Lambda$CDM, hFRW, hFRW+$\Omega_4$ and hFRW+$\Omega_5$ models. }\\
		 %{\color{blue} Is it possible to add one more row of $H_0$ values?}
	 }
 		\begin{tabular} {|c|c|c|c|c|}
 			\hline
 			Parameter & $\Lambda$CDM & hFRW &hFRW$+\Omega_4$ & hFRW$+\Omega_5$ \\
 			\hline
 			{\boldmath$\alpha  $} &-& $0.00032^{+0.00025}_{-0.00020}$& $0.00004^{+0.00065}_{-0.0026}     $&$0.00010^{+0.00012}_{-0.00014}  $\\
 			 			\hline
 			{\boldmath$\Omega_4  $} &-& -& $0.7001\pm 0.0065     $&-\\
 			 			\hline
 			{\boldmath$\Omega_5  $} &-&- & -&$0.4192^{+0.0061}_{-0.0054}  $\\
 			\hline 			
 			{\boldmath$\Omega_m   $} & $0.2960\pm 0.0062       $& $0.3400\pm 0.0070          $& $0.2999\pm 0.0065          $& $0.5108^{+0.0054}_{-0.0061}     $\\
 			\hline 			
 			{\boldmath$\Omega_c h^2   $} & $0.1171\pm 0.0011   $&$0.1067\pm 0.0011          $ &  $0.1178\pm 0.0011   $&$0.1116\pm 0.0012   $    \\
 			\hline 			
 			{\boldmath$100\theta_{MC} $} &$1.0415\pm 0.0007       $& $1.0426\pm 0.0006      $& $1.0410\pm 0.0008      $& $1.0420\pm 0.0007      $\\
 			\hline 			
 			
 			{\boldmath$H_0    $} & $68.85\pm 0.52               $& $60.76\pm 0.38           $& $68.52\pm 0.53           $& $48.19\pm 0.10            $\\
 			
 			$\sigma_8                  $ &  $0.856\pm 0.019     $  & $0.798\pm 0.013      $&$0.800^{+0.020}_{-0.017}$& $0.684^{+0.014}_{-0.013}   $\\
 			\hline 			
 			 $\chi^2                $ &  $3109.66     $  & $3178.46  $&$3102.49$& $3251.54   $\\
 			\hline

 		\end{tabular}
 		\label{tab:2}
 	\end{center}
 \end{table}

In addition to the standard $H_0$ tension, there is another tension between the Planck measurement and the local redshift surveys {(+weak lensing)}. In particular, there is $2\sigma$ tension between the constraints from Planck on the amplitude $\sigma_8$ of the matter fluctuations in linear theory and that from the local measurement. 
We present our MCMC results of the global fitting in Fig.~\ref{fg:6} and Table~\ref{tab:2}. To provide a qualitative analysis, we use the constraints on $S_8= \sigma_8 \sqrt{\Omega_m/0.3}$. From \emph{Planck 2018} with the TTTEEE+lowE dataset~\cite{Aghanim:2018eyx}, we have the $S_8= 0.834 \pm 0.016$. If we take the combination of KV450 and BOSS~\cite{DiValentino:2020vvd} into consideration, the result gives $S_8 = 0.728 \pm 0.026$, which in turn leads to 3.4$\sigma$ tension with Planck data. As is shown in Table~\ref{tab:2}, the best-fit value of $\sigma_8$ in the $\Lambda$CDM model is $\sigma_8=0.856$, which leads to $S_8= 0.850$. As one can see, the result favors a value from the CMB measurement. A similar result can be obtained in the hFRW model, where $\sigma_8$= 0.798 and $S_8= 0.849$. For the hFRW$+\Omega_4$, we obtain a smaller value of $\sigma_8=0.80$,  which gives $S_8= 0.798$. The same is true in the hFRW$+\Omega_5$ model where $\sigma_8 = 0.684^{+0.014}_{-0.013}$ and $S_8=0.6792$; however, the matter energy density in the hFRW$+\Omega_5$ model is surprisingly large, $\Omega_m = 0.5108^{+0.0054}_{-0.0061}$. Thus, based on the values of $\sigma_8$ and $S_8$, we conclude that the $\sigma_8$ tension can be released for the hFRW$+\Omega_4$ model.

\begin{figure}[h!]
\begin{subfigure}{0.4\textwidth}
	\centering
	\includegraphics[width=1 \linewidth, angle=0]{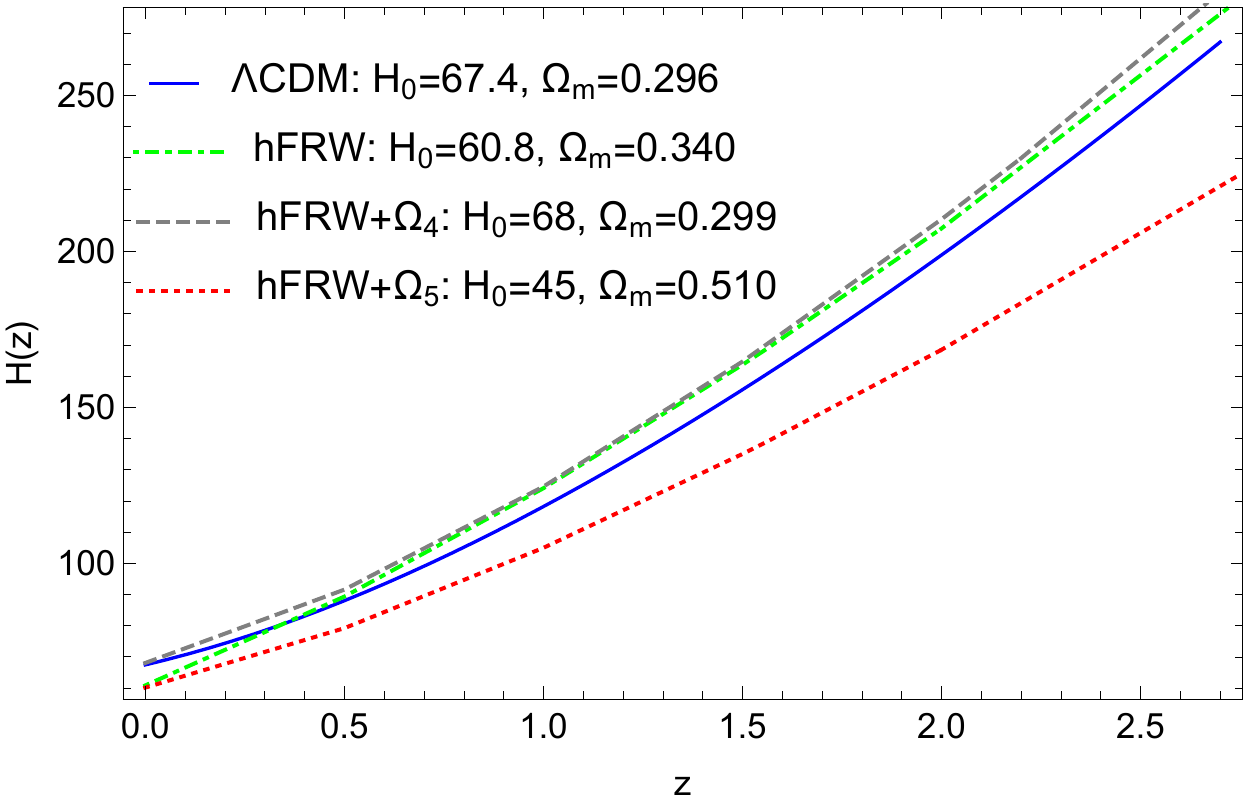}
	\caption{}%The {\color{black} low-redshift, between $0\leq z \leq 3$, evolution of the Hubble parameter $H(z)$.}}	
	\label{fg:7}
\end{subfigure}~~~~\qquad
\begin{subfigure}{0.4\textwidth}
	\centering
	\includegraphics[width=0.98 \linewidth, angle=0]{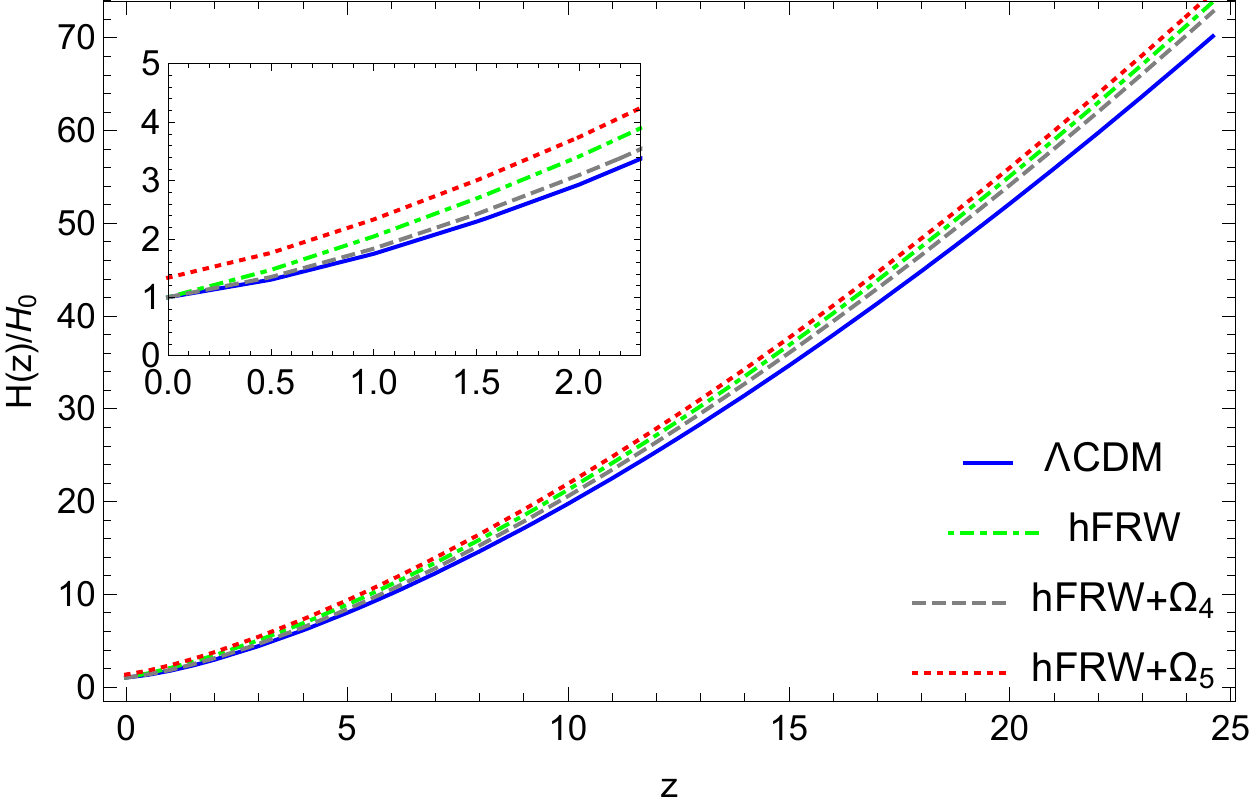}
	\caption{ }%The redshift, between $0\leq z \leq 25$, evolution of the normalized Hubble parameter $H(z)/H_0$.}
	\label{fg:3}
\end{subfigure} %~\\
\caption{\small
The redshift dependence of the Hubble parameter for $\Lambda$CDM, hFRW,  hFRW$+\Omega_4$ and hFRW$+\Omega_5$.}
\label{fig:2}
\end{figure}

In Fig.~\ref{fig:2}, we plot the evolution of the Hubble parameter in terms of the redshift $z$. The dashed gray line in Fig.~\ref{fg:7} indicates that the hFRW$+\Omega_4$ model shows the result closest to the $\Lambda$CDM model, the blue line. Thus, the best-fit value of $H_0=68~\text{km}~\text{s}^{-1}~\text{Mpc}^{-1}$ in the hFRW$+\Omega_4$ model has a preference for Planck data, which is based on the flat $\Lambda$CDM model. The best-fit values of $H_0$ in hFRW and hFRW$+\Omega_5$ models are found to be much smaller than that of the $\Lambda$CDM model, which may be due to larger values of $\Omega_m$ in these models, \emph{i.e.,} the larger the $\Omega_m$ gets, the stronger the gravitational attraction in hFRW and hFRW$+\Omega_5$ models. As is seen in Fig.~\ref{fg:3}, the hFRW, hFRW+$\Omega_4$ and hFRW+$\Omega_5$ models develop a faster expansion rate, $H(z)/H_0$, than that of $\Lambda$CDM result at the same redshift. The holographic models have a faster developing rate than the $\Lambda$CDM model at the same moment. The three holographic models give a shorter history of our Universe from the present to the CMB time, so this is the reason these models cannot release the $H_0$ tension.

% If there was an era in cosmic history of the universe, where $\Omega_I$ gives the most dominant contributions to the total energy density of the universe, $\Omega_I \gg \Omega_M, \Omega_l, \Omega_5$, 

\begin{figure}[h!]
\begin{subfigure}{0.41\textwidth}
	\centering
	\includegraphics[height=0.7\linewidth, width=1.0 \linewidth]{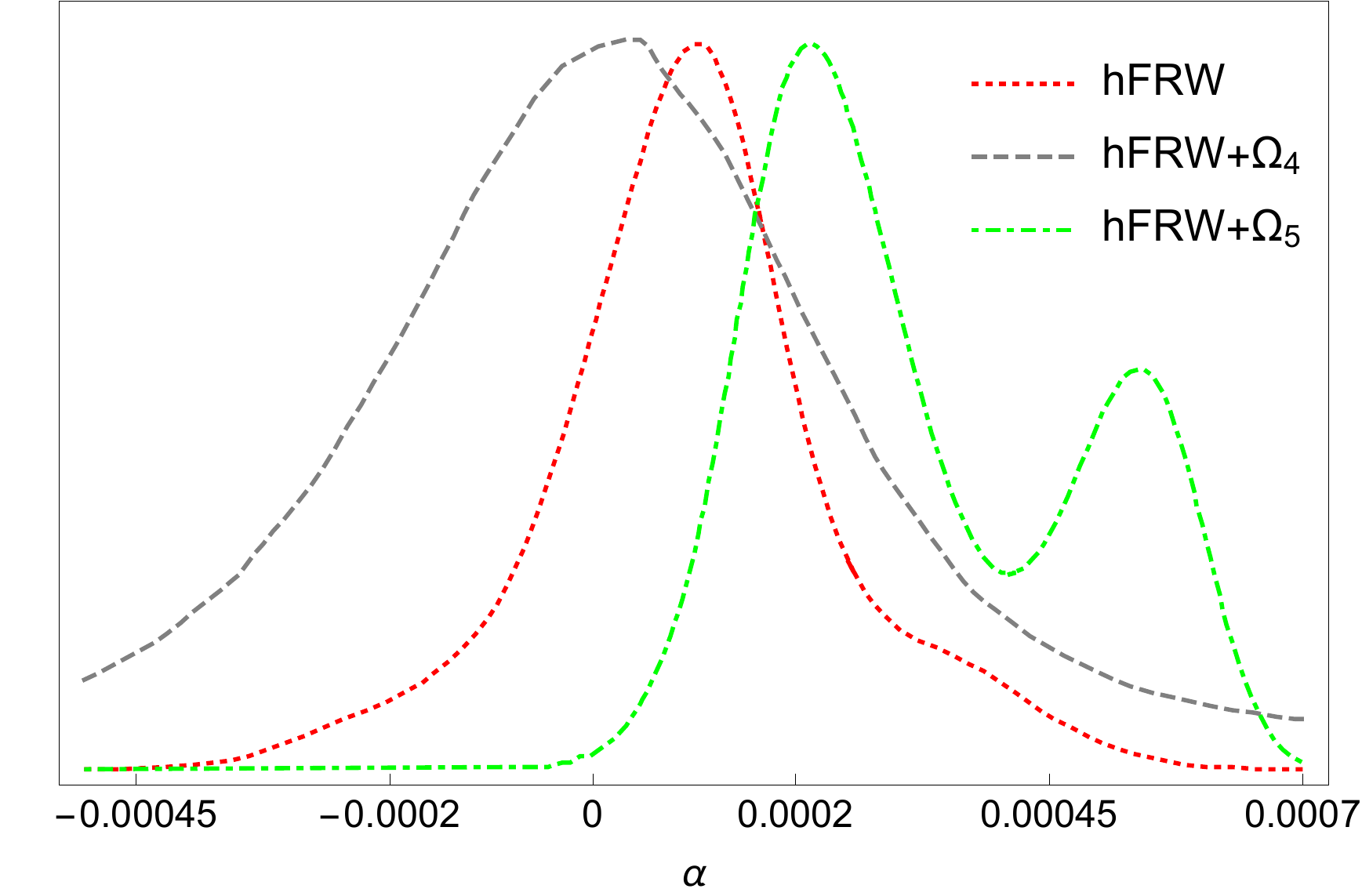}
	\caption{ }
	\label{fg:5}
\end{subfigure}~~~~ 
\begin{subfigure}{0.43\textwidth}
	\centering
	\includegraphics[height=1.1\linewidth, width=0.7 \linewidth, angle=270]{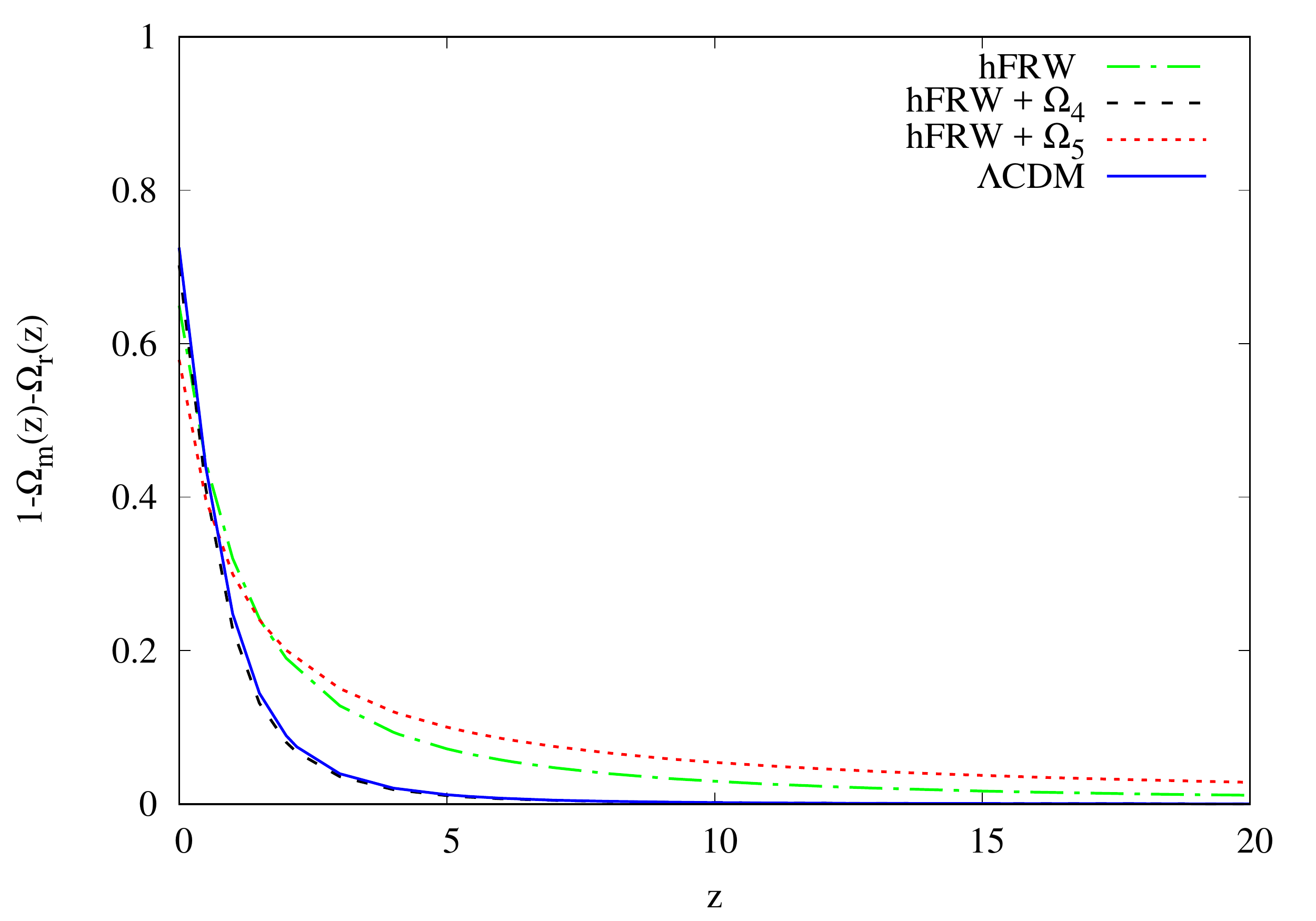}
	\caption{ }
	\label{fg:8}
\end{subfigure} %~\\
\caption{\small (a) One dimensional distributions 95$\%$~C.L. for the $\alpha$ parameter and (b) the redshift dependence of $1-\Omega_m(z)-\Omega_r(z)$ in all $\Lambda$CDM, hFRW, hFRW$+\Omega_4$, and hFRW$+\Omega_5$ models.}\label{fig:3}
\end{figure}

As mentioned earlier, to account for both the negative and the positive contributions of $\Omega_I$, we have chosen the $-1\leq \alpha \leq 1$ as the prior. In Fig.~\ref{fg:5}, therefore, we present the global fitting results on the model parameter $\alpha$. Here, it is worth paying attention to the $\Omega_I(1+z)^4$ quantity under the square root in Eq.~(\ref{E2Eq}). This term can be combined with the contribution of the radiation energy density $\Omega_r$ from $\Omega_M(z)$ under the same square root and be written as $(\Omega_r + \Omega_I)(1+z)^{4}=\tilde{\Omega}_r(1+z)^4$. Since this term is under the square root, it behaves as $\propto \tilde{\Omega}_r(1+z)^2$, giving a curvature like an effect, during the radiation dominated (RD) phase $z > z_{eq}$, where $z_{eq}\sim 1000$ is the redshift at the matter-radiation equality. Then, after the RD phase, its time evolution gets suppressed by a scale factor to some power, and it eventually becomes negligible in the present dark-energy dominated universe. When $\Omega_I \neq 0$, the amount of the additional contribution is $\propto \alpha$ in each holographic model. The case of $\Omega_I = 0$; hence $\alpha=0$, indicates the standard $\Lambda$CDM model. In our numerical result Figure~\ref{fg:5}, the dot-dashed green line corresponds to the hFRW model, and the best-fit value for the model is $\alpha=0.00032^{+0.00025}_{-0.00020}$, see Table~\ref{tab:2}, with the positive center value, which can be well distinguished from zero at $1\sigma$. Although the central value $\alpha= 0.00010^{+0.00012}_{-0.00014}$ in the hFRW$+\Omega_5$ model is still positive, the negative value is possible if the error bar is taken into consideration. A similar result $\alpha=0.00004^{+0.00065}_{-0.0026}$ is obtained in the hFRW$+\Omega_4$ model. As a result, we conclude that the positive energy contribution from $\Omega_I$ is possible in all three holographic models. In contrast, the negative contribution can only be attained in the hFRW$+\Omega_4$ and hFRW$+\Omega_5$ models.

Since we regard the contributions coming from the extra dimension as dark energy, which causes the late-time acceleration of the universe, we show in Fig.~\ref{fg:8} how much each holographic models contribute and how much their contributions would differ from the $\Lambda$CDM model. As the figure shows, the hFRW$+\Omega_4$ model evolves almost as precise as the $\Lambda$CDM model does and gives $ \Omega_D\simeq 0.7$ at $z=0$. The remaining two other models, hFRW and hFRW$+\Omega_5$, show a similar time evolution as the $\Lambda$CDM model, exponentially grow in a redshift decreasing direction. However, compared to the $\Lambda$CDM model, their contributions to the total energy fraction of the universe are found to be less at $z=0$, \emph{e.g.,} $\Omega_{D,\,\text{hFRW}}\simeq 0.66$ and $\Omega_{D,\,\text{hFRW}+\Omega_5}\simeq 0.48$, and more in the earlier time.

\section{Discussions}\label{Sec_Conclusion}
%In this work, we use 3 way to analyze the behavior of Holographic model in 4 and 5 dimensions. 

In summary, we consider the holographic Friedman-Robertson-Walker (hFRW) universe on the (3+1)-dimensional membrane, which is embedded in the (4+1)-dimensional bulk spacetime. The contribution from the extra dimension is represented as the holographic dark fluid on the membrane. In order to fully assess the viability of the hFRW scenario, we derive the corresponding universal modified Friedmann equation by including all of these effects.  In light of the observational data, we have analyzed the observational significance of the hFRW, hFRW+$\Omega_4$, and hFRW+$\Omega_5$ models and provided the best-fitting results of the hFRW, hFRW+$\Omega_4$, and hFRW+$\Omega_5$ models in Table~\ref{tab:2} and Figures.~\ref{fg:6}--\ref{fig:3}. Our numerical analyses have shown that among the three holographic models discussed in this work, the hFRW$+\Omega_4$ model can explain the observational data as elegantly as the $\Lambda$CDM model. Moreover, the hFRW$+\Omega_4$ model can also release the $\sigma_8$ tension effectively. Since the holographic models have a shorter developing history from the present time to the CMB, they lack in solving the $H_0$ tension.  Although the hFRW and hFRW$+\Omega_5$ models are not successful in explaining the observational data or the tensions, they have distinct features to imprint.

Based on our findings, we conclude that among the three holographic models discussed in this work, the hFRW$+\Omega_4$ model can explain the observational data as good as the $\Lambda$CDM model. Moreover, the hFRW$+\Omega_4$ model can also relax the $\sigma_8$ tension. Although the other two models, hFRW and hFRW$+\Omega_5$,  lack explaining the observational data, they have their distinct features to present. Besides the usual hFRW case, where the holographic dark fluid accounts for all the energy density of dark matter and dark energy, we also consider the models with the 4d brane cosmological constant or the 5d bulk one in order to assess the viability of the hFRW scenario fully. For three specific models, including the pure hFRW model, the hFRW model with the brane cosmological constant, and the one with the 5d bulk cosmological constant, we compare theoretical predictions of each model with the observational data using the Markov-Chain-Monte-Carlo sampling method. As a result, it has shown that the parameters in all the considered hFRW models are highly constrained, and only the model with the brane cosmological constant can fit the data as equally well as the standard $\Lambda$CDM model. Interestingly, we find that the $\sigma_8$ tension observed in different large-structure experiments can be effectively relaxed in this hFRW framework.

Finally, we comment on the connection with the generalized holographic cosmology based on the AdS/CFT correspondences \cite{Apostolopoulos:2008ru,Banerjee:2012dw,Camilo:2016kxq,Camilo:2016kxq,Khimphun:2020nkh}, where the holographic stress-energy is living on the boundary of the AdS spacetime. In this paper, we started with the braneworld scenario in AdS. Interestingly, it could be mapped to the moving membrane on the holographic cutoff, where the Brown-York stress-energy tensor on the membrane is identified with that of the holographic dark fluid. 
As has been studied in \cite{Khimphun:2017bac}, if we take the near boundary limit, one can recover the holographic fluid in AdS/CFT correspondences. If we take the near-horizon limit, one can recover the Rindler fluid that was investigated in \cite{Compere:2011dx,Compere:2012mt,Eling:2012ni,Cai:2013uye,Cai:2014ywa,Pinzani-Fokeeva:2014cka}, which is a new perspective on the membrane paradigm of black holes. 
%A similar scenario has also been investigated in Refs.~\cite{Pourhasan:2013mqa,Altamirano:2017wqm}.
Actually, a similar scenario has also been discussed  in Refs.~\cite{Pourhasan:2013mqa,Altamirano:2017wqm} under the name of the holographic big bang model, in which the early universe on the (3+1)-dimensional brane arise out of a collapsing star in the (4+1)-dimensional bulk. Their motivation was to avoid the classical Big-Bang singularity and to generate the scale-invariant cosmological perturbations from the brane atmosphere around the bulk black hole horizon. In comparison, the hFRW model in Ref.~\cite{Cai:2018ebs} focused on the nature of the dark sector and its effects on the late-time evolution of the universe. It will be interested to make further detailed studies based on this frame work. 

%where the Brown-York stress-energy tensor is used. . 

%In order to fully account for the phenomenology of this scenario, 
%When taking the near-horizon limit, one could reach the so-called Rindler fluid 

\section*{Acknowledgments}
\label{Acknowledgment}\small 
D. Huang and Y. L. Zhang are supported by National Natural Science Foundation of China (NSFC) under grants No. 12005254, 12005255.
B. H. Lee, G. Tumurtushaa and L. Yin are supported by Basic Science Research Program through the National Research Foundation of Korea (NRF) funded by the Ministry of Education through the Center for Quantum Spacetime (CQUeST) of Sogang University 
(NRF-2020R1A6A1A03047877, NRF-2020R1F1A1075472).
G. Tumurtushaa is also supported by Ministry of Science and Technology (MoST) under grant No. 109-2112-M-002-019. We thank many helpful conversations with R. G. Cai, S.~Mukohyama, M. Sasaki and S. Sun on this topic.

\appendix

\section{FRW universe on the brane world scenario}
%{Solving the Einstein equations in Brane world coordinate}
\label{App1}
In this appendix, we show how to obtain the functions $A(w,t)$ and $N(w,t)$ defined in the Gaussian normal metric in Eq.~(\ref{Metric5D}) by solving the bulk Einstein equation
\begin{eqnarray}\label{Einstein5d}
\tilde{G}_{MN} +\Lambda_5 \tilde{g}_{MN} = 0\,.
\end{eqnarray}
As a result, the non-trivial equations in the bulk are listed as follows:
\begin{itemize}
\item $(tw)$-component:
\begin{equation}\label{Eqtw}
\frac{3}{aAN}\left[N^\prime (A\dot{a}+a\dot{A})-N(\dot{a}A^\prime + a \dot{A}^\prime)\right] = 0\,,
\end{equation}
\item $(tt)$-component:
\begin{eqnarray}\label{Eqtt}
3\left[H^2 + 2 H \frac{\dot{A}}{A}-N^2\left(\frac{A^{\prime \, 2}}{A^2}+\frac{A^{\prime\prime}}{A}\right)+\frac{\dot{A}^2}{A^2}\right]-\Lambda_5 N^2 =0\,,
\end{eqnarray}
\item $(ww)$-component:
\begin{eqnarray}\label{Eqww}
\frac{3}{a^2 A^2 N^3} &\bigg\{& a^2 N (N^2 A^{\prime\,2}-\dot{A}^2)-A^2\left[N(\dot{a}^2+a\ddot{a})-a\dot{a}\dot{N}\right]\nonumber\\
&+& aN\left[-4N\dot{a}\dot{A}+a(N^2 A^\prime N^\prime+\dot{A}\dot{N}-N\ddot{A})\right]\bigg\}+\Lambda_5 = 0\,,
\end{eqnarray}
\item $(ii)$-component:
\begin{eqnarray}\label{Eqii}
\frac{1}{N^3} &\bigg\{& a^2 N(N^2 A^{\prime\,2}-\dot{A}^2) + A^2 \left[-N(\dot{a}^2+2a\ddot{a})+a^2 N^2 N^{\prime\prime}+2a\dot{a}\dot{N}\right] \nonumber\\
&& +2 aA\left[-3N\dot{a}\dot{A} + a(N^2 A^\prime N^\prime + N^3 A^{\prime\prime}+\dot{A}\dot{N}-N\ddot{A})\right] \bigg\} + \Lambda_5 a^2 A^2 = 0\,.
\end{eqnarray}
\end{itemize}
From Eq.~(\ref{Eqtw}), we have
\begin{eqnarray}
\left(\ln N\right)^\prime = \left[\ln (\dot{a}A + a \dot{A})\right]^\prime\,.
\end{eqnarray}
With the boundary conditions in Eq.~(\ref{bc0}), the integration over $w$ gives us
\begin{eqnarray}\label{EqN}
N= \frac{\partial_t [aA]}{\dot{a}} = A + \dot{A}/H\,.
\end{eqnarray}
Furthermore, with the relation between $N$ and $A$ in Eq.~(\ref{EqN}), the $(tt)$-component Einstein equation in Eq.~(\ref{Eqtt}) can be transformed into
\begin{equation}
(A^2)^{\prime\prime} -\lambda_5^2 A^2 = 2H^2\,,
\end{equation}
where $\lambda_5^2 \equiv -2\Lambda_5/3$. The most general solution to this equation is given by
\begin{eqnarray}
A^2 = -\frac{2 H^2}{\lambda_5^2} + C_1 \cosh(\lambda_5 w) + C_2 \sinh(\lambda_5 w)\,.
\end{eqnarray}
By using the boundary condition $A^2(w=0) = 1$, we can determine $C_1 = 1+ 2H^2/\lambda_5^2$
\begin{eqnarray}
A^2 = \cosh{(\lambda_5 w)} - \frac{2H^2}{\lambda_5^2} \left[1- \cosh(\lambda_5 w)\right] + C_2 \sinh(\lambda_5 w)\,.
\end{eqnarray}
In order to determine $C_2$, we firstly note that $C_2$ can be a function of the time coordinate $t$ without any dependence on $w$. Thus, we can solve $C_2$ just on the brane where $w=0$. Note that on the brane, the $(ww)$-component equation in Eq.~(\ref{Eqww}) is simplified into
\begin{eqnarray}
A^{\prime\,2} - \left(H^2 + \frac{\ddot{a}}{a}\right) A^2 + \frac{N^\prime}{N}A^{\prime}A = \frac{1}{2}\lambda_5^2 A^2\,. 
\end{eqnarray}
When written in terms of the function $C_2$, this equation is given by
\begin{eqnarray}
\frac{1}{4H} \frac{dC_2^2}{dt} + (C_2^2 -1) = \frac{2}{\lambda_5^2} (2H^2 + \dot{H})\,,
\end{eqnarray}
which can be solved to be
\begin{eqnarray}
C_2^2 = 1 + \frac{4H^2}{\lambda_5^2} + \frac{C_0}{a^4}\,,
\end{eqnarray}
where $C_0$ is a constant of integration. Therefore, 
\begin{eqnarray}\label{BulkSol}
A^2 = \cosh(\lambda_5 w) - \frac{2H^2}{\lambda_5^2} \left[1-\cosh(\lambda_5 w)\right] + \sigma \frac{2\sinh(\lambda_5 w)}{\lambda_5} \sqrt{\frac{\lambda_5^2}{4} + H^2 + \frac{I}{a^4}} \,,
\end{eqnarray}
with $I\equiv \lambda_5^2 C_0/4$ and $\sigma= \pm 1$. 
%By taking the limit $\Lambda_5 \to 0$, the solution above can be further simplified to
%\begin{eqnarray}
%A^2 = 1 - H^2 w^2 + 2\sigma  w\sqrt{H^2 + I/a^4}\,.
%\end{eqnarray}
Finally, note that the $(ii)$-component bulk Einstein equation is not used in this derivation since it cannot give any additional information when confined onto the brane.

%\section{Detailed derivation}
\section{FRW universe on the holographic cutoff}\label{App2}
 % (for the 2nd ones)}
In this appendix, we present some calculation details for the second derivation of the modified Friedman equations given in Sec.~\ref{SecMovingBrane}. 
Here we give a derivation of the AdS black brane metric, the most general solution to the bulk Einstein equation in Eq.~(\ref{EinsteinBulk}). Taking following ansatz for the bulk metric tensor:
\begin{eqnarray}
d\tilde{s}^2_5 = -A(r)d{\tc}^2 + B(r)dr^2 + \frac{r^2}{L^2} dx_i^2\,, 
\end{eqnarray}
where we defined the AdS length through $L^2\equiv -6/\Lambda_5$. Following the bulk Einstein equation in Eq.~(\ref{EinsteinBulk})
we can derive the following non-trivial differential equations for the functions $A(r)$ and $B(r)$:
\begin{itemize}
\item $(t,t)$-component: 
\begin{eqnarray}\label{5dtt}
\frac{3A(-B+rB^\prime)}{2r^2 B^2} - \Lambda_5 A = 0\,,
\end{eqnarray}
\item $(r,r)$-component: 
\begin{eqnarray}\label{5drr}
\frac{3(2A+rA^\prime)}{2r^2 A} + \Lambda_5 B = 0\,.
\end{eqnarray}
\end{itemize}
From Eq.~(\ref{5dtt}), we can obtain the following solution to the function $B(r)$ 
\begin{eqnarray}
B(r) = \left(\frac{r^2}{L^2}-\frac{m}{r^2}\right)^{-1}\,,
\end{eqnarray}
where $M$ is the integration constant. By putting the above solution of $B(r)$ into Eq.~(\ref{5drr}) and integrating over $r$, we can yield the function of $A(r)$ as follows
\begin{eqnarray}
A(r) = C_1 \left(\frac{r^2}{L^2} - \frac{m}{r^2}\right)\,,
\end{eqnarray}
in which $C_1$ is another integration constant. However, if we rescale the coordinate $t \to t/\sqrt{C_1}$, we can absorb this constant into the definition of the metric. As a result, the final solution to the bulk Einstein equation with a cosmological constant is given by Eq.~(\ref{MetricBulk}) in which $M$ can be identified with the black brane's mass. 
% with $C$ identified with $M$ in Yun-Long's notation. 
%\subsection{Calculation on the extrinsic curvature}\label{App2Ext}

Here we show a derivation of the expressions in Eq.~(\ref{ExtK3}) for the extrinsic curvature of the brane, for which we shall strictly follow its definition ${\cal K}_{\mu\nu} = e^{M}_{(\mu)} e^N_{(\nu)} \nabla_M n_N$. Let us begin by listing all of the nonzero components of the bulk connection for the metric in Eq.~(\ref{MetricBulk}):
\begin{eqnarray}
&&\Gamma^{{\tc}}_{{\tc}r} = \Gamma^{\tc}_{r{\tc}} = \frac{f^\prime}{2f}\,, \quad \Gamma^r_{{\tc}{\tc}} = \frac{f f^\prime}{2}\,,\quad \Gamma^r_{rr} = -\frac{f^\prime}{2f} \,,\nonumber\\
&& \Gamma^r_{ii} = -\frac{r f}{L^2} \,,\quad \Gamma^i_{ir} = \Gamma^i_{ri} = \frac{1}{r}\,.
\end{eqnarray}
Then it is straightforward to obtain the $(ii)$ components of the extrinsic curvature:
\begin{eqnarray}
{\cal K}^+_{ii} = \nabla_i n_i = -\Gamma^r_{ii} n_r = \frac{\sigma a}{L} \sqrt{f+L^2 \dot{a}^2}\,.
\end{eqnarray} 
However, in order to obtain the correct result of the $({\tta}{\tta})$ component, it is useful to remark that the metric in Eq.~(\ref{MetricBulk}) only depends on the coordinate $r$, without any reliance on the bulk time coordinate $t$. Therefore, the time derivative of any component in the metric tensor and in the normal vector $n_M$ should vanish, {\it i.e.,} $\partial_t n_M = 0$. Under this caveat, we can compute ${\cal K}^+_{{\tta}{\tta}}$ as follows:
\begin{eqnarray}
{\cal K}^+_{{\tta}{\tta}} &=& e^{\tc}_{({\tta})} e^{\tc}_{({\tta})} \nabla_{\tc} n_{\tc} + e^{\tc}_{({\tta})} e^r_{({\tta})} \nabla_{\tc} n_r + e^r_{({\tta})} e^{\tc}_{({\tta})} \nabla_r n_{\tc} + e^r_{({\tta})} e^r_{({\tta})} \nabla_r n_r \nonumber\\
&=& - e^{\tc}_{({\tta})} e^{\tc}_{({\tta})} \Gamma^r_{{\tc}{\tc}} n_r - e^{\tc}_{({\tta})} e^r_{({\tta})} \Gamma^{\tc}_{{\tc}r} n_{\tc} + e^r_{({\tta})} e^{\tc}_{({\tta})} \nabla_r n_{\tc}+ e^r_{({\tta})} e^r_{({\tta})} \nabla_r n_r \nonumber\\
&=& e^{\tc}_{({\tta})} \partial_{\tta} n_{\tc} + e^r_{({\tta})} \partial_{\tta} n_r - e^{\tc}_{({\tta})} e^{\tc}_{({\tta})} \Gamma^r_{{\tc}{\tc}} n_r - 2 e^{\tc}_{({\tta})} e^r_{({\tta})} \Gamma^{\tc}_{{\tc}r} n_{\tc} - e^r_{({\tta})} e^r_{({\tta})} \Gamma^r_{rr} n_r\,.
\end{eqnarray}
By putting nonzero components of the normal vector and the connection into the above formula, we can obtain the $({\tta}{\tta})$-component of the extrinsic curvature as in Eq.~(\ref{ExtK3}).

\end{document}